\documentclass[letterpaper,11pt]{article}
\pdfoutput=1

\usepackage{jheppub}
\usepackage{mathtools,amssymb,braket,mathrsfs,tikz,subfig,xspace,xcolor,float,color,siunitx,comment,afterpage,multirow,array,hhline,amsmath,amsthm,framed,colortbl, bm}
\usepackage[countmax]{subfloat}

\newcommand{\be}{\begin{equation}}
\newcommand{\ee}{\end{equation}}

\newcommand{\E}{\mathcal{E}}

\usepackage{amsmath,amssymb,amsthm,cancel,hyperref,graphicx,xcolor}
\usepackage{mathrsfs,wasysym}
\usepackage{booktabs}
\usepackage{tikz}
\usetikzlibrary{calc}
\usepackage{placeins}

\usepackage{slashed}
\usepackage{graphicx}

\usepackage{soul}

\usepackage{array}

\usepackage{comment}

\usepackage{url}
\usepackage{xspace}
\usepackage{dsfont}
\usepackage{amssymb}
\usepackage{amsmath}
\usepackage{graphicx}
\usepackage{hhline}
\usepackage{physics}
\usepackage{color}
\definecolor{mit-red}{rgb}{0.64,.12,0.2}
\definecolor{darkred}{rgb}{1.0,0.1,0.1}
\definecolor{darkgreen}{rgb}{0.1,0.7,0.1}
\definecolor{darkblue}{rgb}{0.1,0.1,1.0}

\usepackage{amsmath}

\DeclareRobustCommand{\Sec}[1]{Sec.~\ref{sec:#1}}

\DeclareRobustCommand{\Tab}[1]{Table~\ref{tab:#1}}

\DeclareRobustCommand{\Fig}[1]{Fig.~\ref{fig:#1}}
\DeclareRobustCommand{\Figs}[2]{Figs.~\ref{fig:#1} and \ref{fig:#2}}

\def\hepph{\texttt{hep-ph}}
\def\hepex{\texttt{hep-ex}}

\def\arxiv{\texttt{arXiv}}

\begin{document}

% Magic to fix footnote behavior.
\count\footins = 1000
\interfootnotelinepenalty=10000
\setlength{\footnotesep}{0.6\baselineskip}

%Change tables
\renewcommand{\arraystretch}{1.3}

\title{A Search for ``New Physics'' ``Beyond the Standard Model'' in Open Data with Machine Learning\\ {\small April 1, 2025}}

\author[a,b]{Rikab Gambhir,}

\affiliation[a]{Center for Theoretical Physics, Massachusetts Institute of Technology,  Cambridge, MA 02139, USA}
\affiliation[b]{The NSF AI Institute for Artificial Intelligence and Fundamental Interactions, USA}

\emailAdd{rikab@mit.edu}

% \preprint{MIT-CTP 5771}

\abstract{
In this new era of large data, it is important to make sure we do not miss any signs of new physics.
Using the publicly-available open data collected by the arXiv.org experiment in the \texttt{hep-ph} channel, corresponding to a raw total integrated $\mathcal{L}$iterature of 65,276 papers, we perform a search for ``New Physics'' and related signals.
In the worst-case, we are able to detect ``New Physics'' with ``the LHC'' at a significance level of at least $6.5\sigma$.
This ``New Physics'' signature is primarily ``Dark'' in nature, and is potentially axion(-like) dark matter.
We also show the potential for further improvement in the future, and that ``New Physics'' can be found with ``a Future Collider'' at at least $8.9\sigma$, as well as the potential to find ``New Physics'' without any collider at all.
This search is performed using code that was $80\%$ written by Machine Learning methods.
}

\maketitle

\section{Introduction}

Using tools like the Large Hadron Collider (LHC), we have made enormous progress in our understanding of the Standard Model (SM) over the last few decades.
However, little progress has been made in finding new physics Beyond the Standard Model (BSM)~\cite{Peskin:1997ez}, despite indirect evidence of its existence.
In recent years the paradigm of \emph{anomaly detection}~\cite{hepmllivingreview, ATLAS:2020iwa, ATLAS:2023azi, CMS:2024nsz, ATLAS:2023ixc, ATLAS:2025obc, knapp2020adversarially, Gambhir:2025afb} has grown in popularity: new physics might be buried in data we already have, if only we had a method to uncover it.

In this paper, we perform a direct search for ``New Physics'', using 126 GB of data from \url{https://arXiv.org/archive/hep-ph}, collected from January 1$^{\rm st}$, 2016 to March 1$^{\rm st}$ 2025 --- our preprocessed dataset and analysis framework will be made public (see \Sec{code_and_data}).
By analyzing the abstracts of randomized \texttt{arXiv} submissions in the \texttt{hep-ph} channel, we are able to find \emph{significant} excess of ``New Physics'', most of which is ``Dark'', well above the 5$\sigma$ discovery threshold.
We emphasize that this is in \emph{already existing} data produced by the physics community over the last decade --- no new experiment is needed.\footnote{Despite some great recommendations for experiments~\cite{Cesarotti:2023sje}.}
Moreover, our analysis strategy is multi-differential: 
We are able to gauge the impact of experiments by searching for ``New Physics'' with ``the LHC'' or with ``Future Colliders'' -- the later of which we will see has a higher ``New Physics'' significance.
Lastly, our technique can be used to analyze trends in the particle physics community over time.
We will see, for example, that ``the LHC'' is not as popular as it once was.

A major component of this search is the use of Machine Learning (ML)-based techniques.
Unlike traditional ML-based searches, however, which typically employ ML to perform fits to data~\cite{Collins:2019jip}, we instead employ ML to write most of the analysis framework.
We show that it is possible to use ChatGPT to code about $80\%$ of a novel and gimmicky data analysis in less than 18 hours, including sleep.\footnote{Only the analysis code was generated with the help of AI, not the text of this paper or the plots. I have too much pride to let an AI speak for me. Machines have no right to utter humanity's sacred tongues.}
We hope this serves as a lesson to the particle physics community --- AI can help one realize an idea quickly, but it cannot replace the soul behind it.

The rest of this paper is organized as follows:
In \Sec{data}, we describe the \arxiv\, dataset and our selection cuts.
In \Sec{results}, we show and discuss our main results.
Finally, we present our outlook in \Sec{conclusions}.

\section{The Dataset and Methodology}\label{sec:data}

Using the extremely convenient \texttt{arxiv} (v2.1.3) Python Package~\cite{arxiv_scraper},
it is possible to analyze \texttt{arXiv} data.
However, because the full \texttt{arXiv} dataset is extremely large, and most of this data does not contain ``New Physics'' or related signals, it is wise to make selection cuts.
I am also limited by the fact that my laptop only has 8 GB of RAM and that I really have to get this out by April $1^{\rm st}$.

We will search in the \hepph\, channel, since it is unknown how likely it is to find ``New Physics'' in other channels such as \texttt{cond-mat} or the quantitative finance categories.
We will use \texttt{hep-ex} as a control group, since supposedly no collider experiment has ever seen ``New Physics''.
We will restrict ourselves from January 2016 to March 2025, a 9-year data taking period.
The total integrated $\mathcal{L}$iterature\footnote{We use $\mathcal{L}$iterature rather than the more commonly-used \emph{Luminosity} because not all papers are equally illuminating.} collected during this period, $\int dt\, \mathcal{L}$, is
\begin{align}
    \hepph: \int dt\, \mathcal{L} = 65,276 \text{ Papers}. \\
    \texttt{hep-ex}: \int dt\, \mathcal{L} = 25,661 \text{ Papers}.
\end{align}
An average paper is approximately 1.93 MB, corresponding to approximately 126 GB of collected data.
In its raw AOD (All Of the Data) file format, the data is too unwieldy for phenomenological analysis and must be preprocessed.

\begin{figure}
    \centering
    \includegraphics[width=0.95\linewidth]{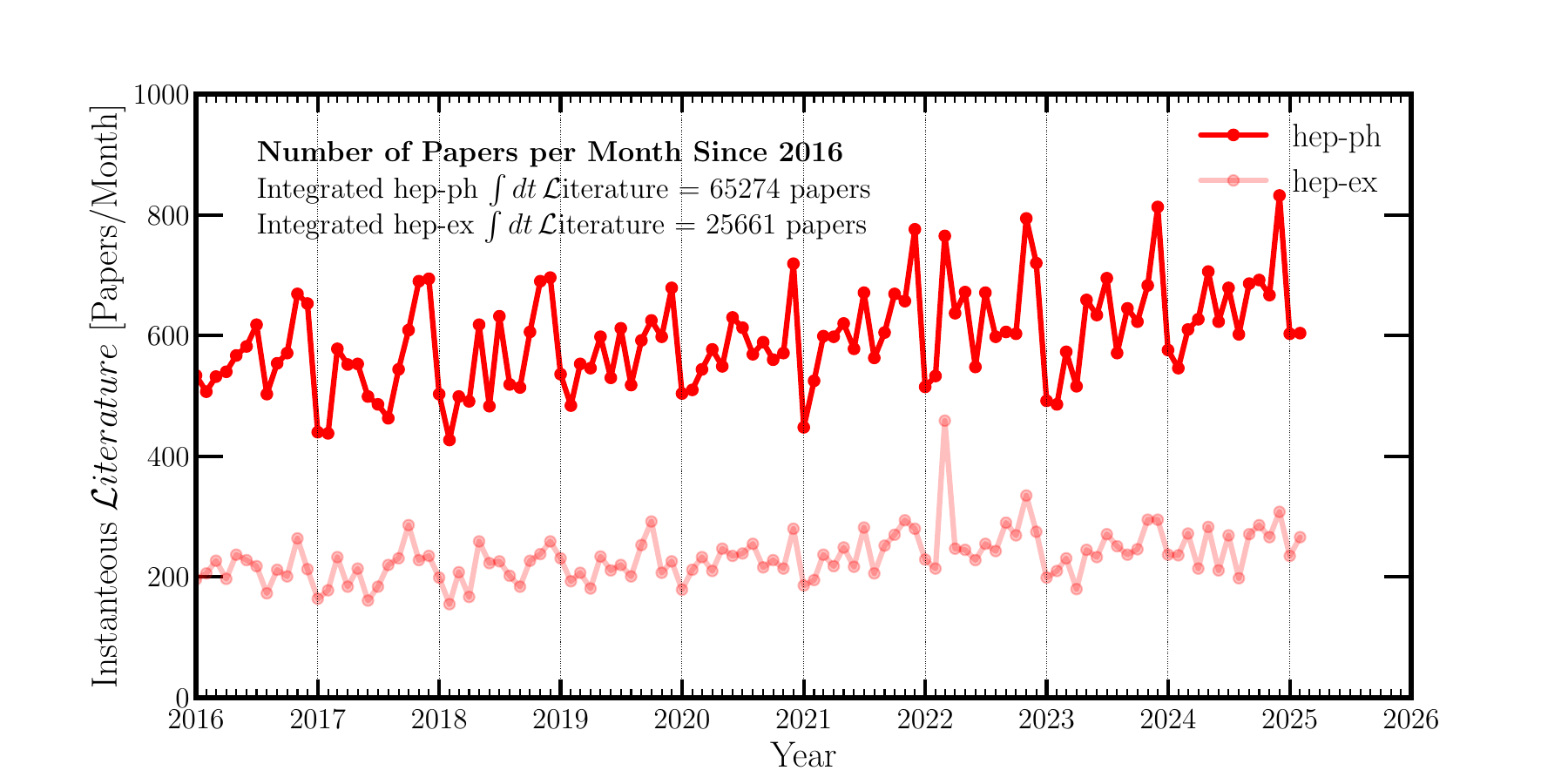}
    \caption{The instantaneous $\mathcal{L}$iterature collected over time for the \hepph\, and \texttt{hep-ex} channels.}
    \label{fig:literature}
\end{figure}

We show a plot of the instantaneous $\mathcal{L}$iterature rate, measured per-month, in \Fig{literature}.
It is interesting to note in \Fig{literature} the bump before the large dip at the end of each year.
This happens to coincide with a number of annual astrophysical phenomenon:
\begin{itemize}
    \item Earth reaches its Perihelion in its orbit around the Sun.
    \item The Quadrantid meteor shower.
    \item Orion becomes visible.
\end{itemize}
This happens 9 times in a row and is unlikely to be a coincidence --- 18 times in a row if $\hepex$ is included, though it seems to happen at a lesser extent there.
We leave the study of this phenomenon for potential future work, which will likely involve a similar analysis in the \texttt{astro-ph}\, channel.

\subsection{Selection Cuts}

Due to the extremely high data rate, our selection must be prescaled~\cite{Smith:2016vcs}.
The data collection trigger only fires exactly 250 times per month, heavily reducing our effective $\mathcal{L}$iterature.

Each \arxiv\, entry is a composite combination of many different sections (a title, an abstract, sections, a conclusion, and so on), distributed in a Portable Document Format (PDF)~\cite{Placakyte:2011az} file.
In a typical PDF, the fundamental abstract field will only be a small fraction ($1-10\%$) of the total document, and the rest of the document typically contains no additional interesting physics. 
We make a hard cut to only select the abstract and remove the rest of the Underlying Entry (UE).
After these two cuts, the effective data size has been reduced from 126 GB to 34 MB.
We emphasize that we will make all of this curated data public for further analysis by the community.

Next, we can perform our search. 
For any entry $\E$, its abstract is a character string $A_\E$.
We would like our analysis to be IRC (Improved-Readability-and-Capitalization)-safe, so we cut on only Secondary Verbiage (SV) filler words: ``the'', ``of'', ``for'', ``a'', ``an''; leaving behind only Primary Verbiage (PV) from the main information content.
We remove all punctuation and spaces and everything is made lowercase.
Trailing ``s'''s are removed from the end of words so that everything is singular.
This ensures our observables are robust with respect to collinear splittings (``standardmodel'' $\to$ ``standard model'') and infrared emission (``standardmodel'' $\to$ ``the standardmodel'')\footnote{This can be taken further, in fact.
The Hamming Distance on strings encodes this notion of IRC-safety topologically (a string is robust against edits), which means you can actually define metric-based observables of the type explored in ~\cite{Komiske:2020qhg, Cesarotti:2020hwb, Ba:2023hix, Gambhir:2024ndc} using this. We will not explore this here, but somebody should.}
Given a  \emph{keyword} character string $x$, such as ``New Physics'', we also define its IRC-safe projection, e.g. ``newphysic''.
Then, we can define:
\begin{align}
    O_x(\E) = \text{Number of times } x \text{ appears in } A_\E 
\end{align}
For example, for $\E = $\texttt{A Search for ``New Physics'' ``Beyond the Standard
Model'' in Open Data with Machine Learning}, where $x = $``New Physics'', we have $O_x(\E) = 4$.

We will find it useful to define \emph{collections} of keywords, rather than wording with individual ones.
This is to capture related concepts: it would be unfair if an entry found ``Supersymmetry'', but it was not included in our selection.
In a sense, we are Looking Elsewhere to improve our global odds of finding \emph{any} ``Signal'', which is very well-motivated statistically.
The full list of keyword selections is shown in \Tab{keyword_categories}.

\begin{table}[ht]
\centering
\begin{tabular}{|l|p{10cm}|}
\hline
\textbf{Category} & \textbf{Keywords} \\ \hline
``New Physics'' & Beyond the Standard Model, BSM, Signal, New Physics, Anomaly, Exotic, SUSY, Supersymmetry, Axion, ALP, Dark \\ \hline
``the LHC'' & CMS~\cite{CMS:2008xjf}, ATLAS~\cite{ATLAS:2008xda}, LHC, CERN, Fermilab, Large Hadron Collider \\ \hline
``Future Colliders'' & FCC~\cite{Benedikt:2020ejr}, Future, Muon collider~\cite{AlAli:2021let}, Linear collider, CLIC~\cite{Brunner:2022usy}, ILC~\cite{Battaglia:2007rk}, muC \\ \hline
``Background'' & Background, Noise, Pileup, Standard Model, SM \\ \hline
\end{tabular}
\caption{Keyword Categories}
\label{tab:keyword_categories}
\end{table}

Given this, we can plot the search results over time for each category and keyword, and look for trends.
We may also count the number of ``New Physics'' events we see, $S$.
To estimate the significance of the ``New Physics'', we need a background estimate.
Without any additional physics insight or cuts, the naive background estimate $B_{\rm naive}$ is simply the total number of words in the event.
However, we can make cuts to remove some of this background.
Instead, we can look for the ``Background'' keyword (and its related keywords, defined in \Tab{keyword_categories}, and cut everything else away.
This severely reduces the background with absolutely zero ``Signal'' loss, which not even ML-based anomaly detection methods can do.
The total count of ``Background'' events is $B_{\rm less\,naive}$.

After all of the above cuts, assuming each event is a Poisson random variable (meaning there is a nonzero chance for ``New Physics New Physics'' or ``bsmbsmbsmbsmbsm'' to appear in text),  we can estimate the discovery significance:
\begin{align}
    \sigma = \frac{S}{\sqrt{B}}.
\end{align}
Moreover, we can condition both $S$ and $B$ on other keywords: we only count an event towards $S$ or $B$ if it contains the conditional keyword. 
Here, in addition to the inclusive search, we will perform a differential study exclusive in the collider type: finding ``New Physics'' with ``the LHC'', and finding ``New Physics'' with a ``Future Collider''.

\section{Main Results}\label{sec:results}

\begin{figure}
    \centering
    \includegraphics[width=0.99\linewidth]{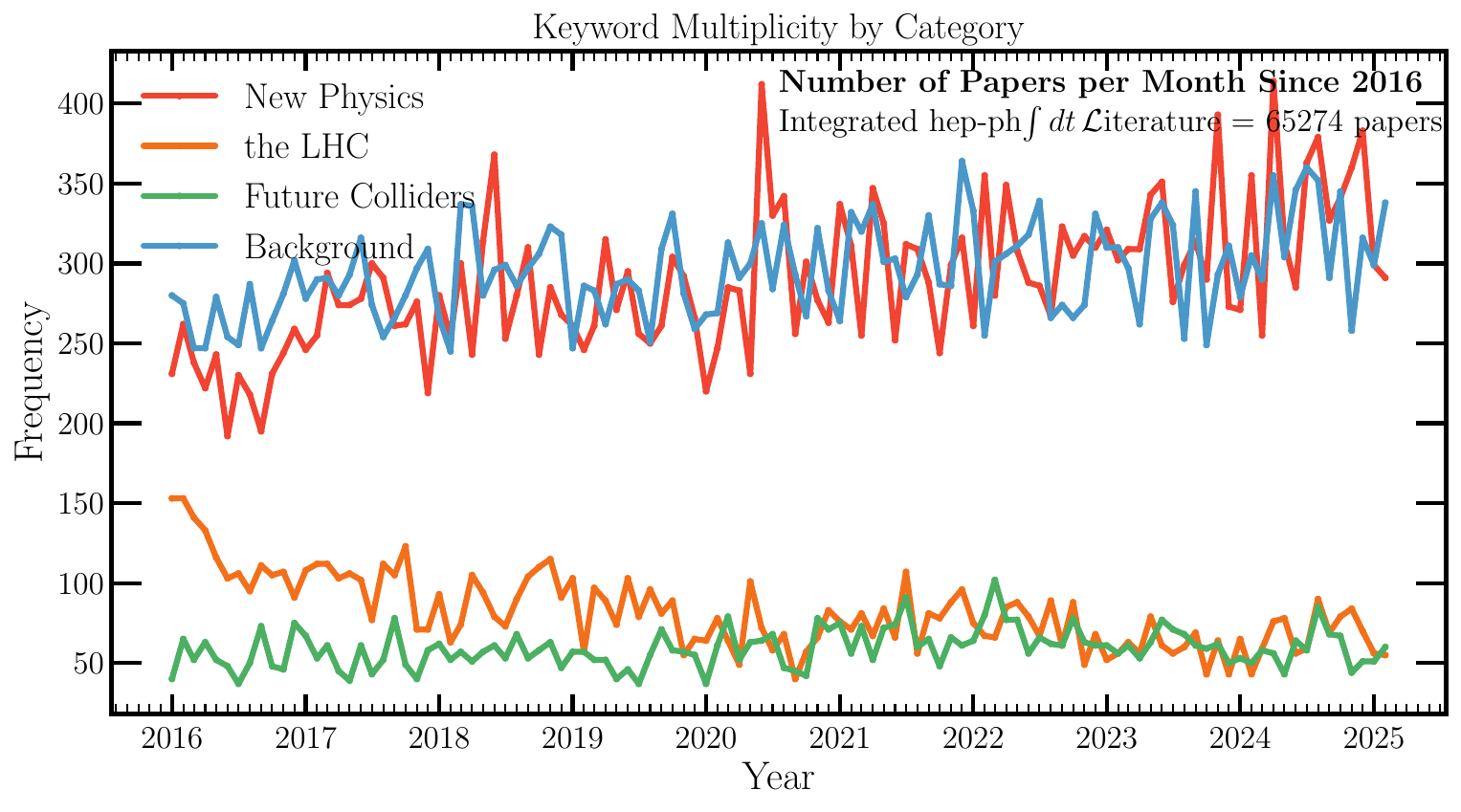}
    \caption{Total event count per month of the categories defined in \Tab{keyword_categories} for \hepph.}
    \label{fig:hep-ph-categories}
\end{figure}

In \Fig{hep-ph-categories}, we plot the total event count for each keyword category (as defined in \Tab{keyword_categories}), binned per month.
We immediately discern several striking features:
\begin{enumerate}
    \item \textbf{More ``New Physics''}: The frequency of ``New Physics'' has been steadily increasing over time. More and more people are talking about new things! Of course, the ``Background'' is just as prevalent.
    \item \textbf{Less ``LHC''}: The ``LHC'' has been declining in popularity over time. This suggests that the excitement for Runs 4-10 of the LHC, HL-LHC, and LH-HL-LHC do not have enthusiastic support in the community, as they have lost interesting in finding nothing.
    \item \textbf{The Future hasn't changed}: The discussion of ``Future Colliders'' has been relatively constant for about a decade. %
    This suggests that physicists' interest in future projects is entirely uncorrelated with whether such projects ever actually get built.
\end{enumerate}

\begin{figure}[tbp]
    \centering
    \subfloat[New Physics]{%
        \includegraphics[width=0.99\linewidth]{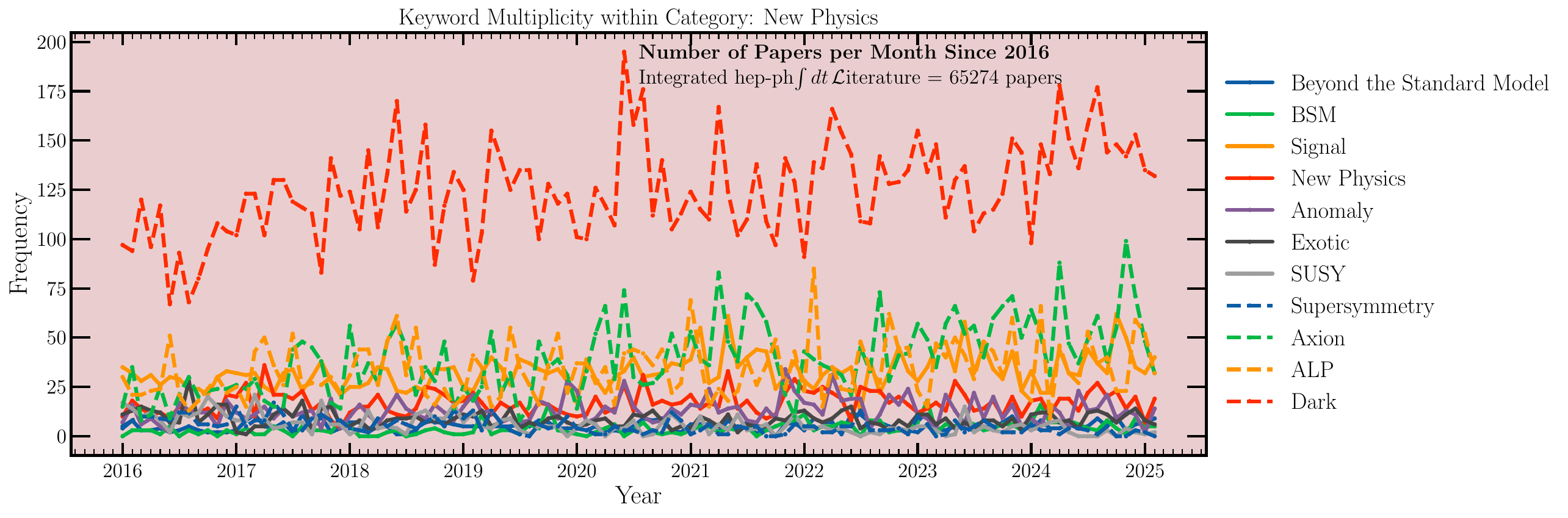}%
        \label{fig:newphysics}%
    }\\[0.1cm]
    \subfloat[Background]{%
        \includegraphics[width=0.99\linewidth]{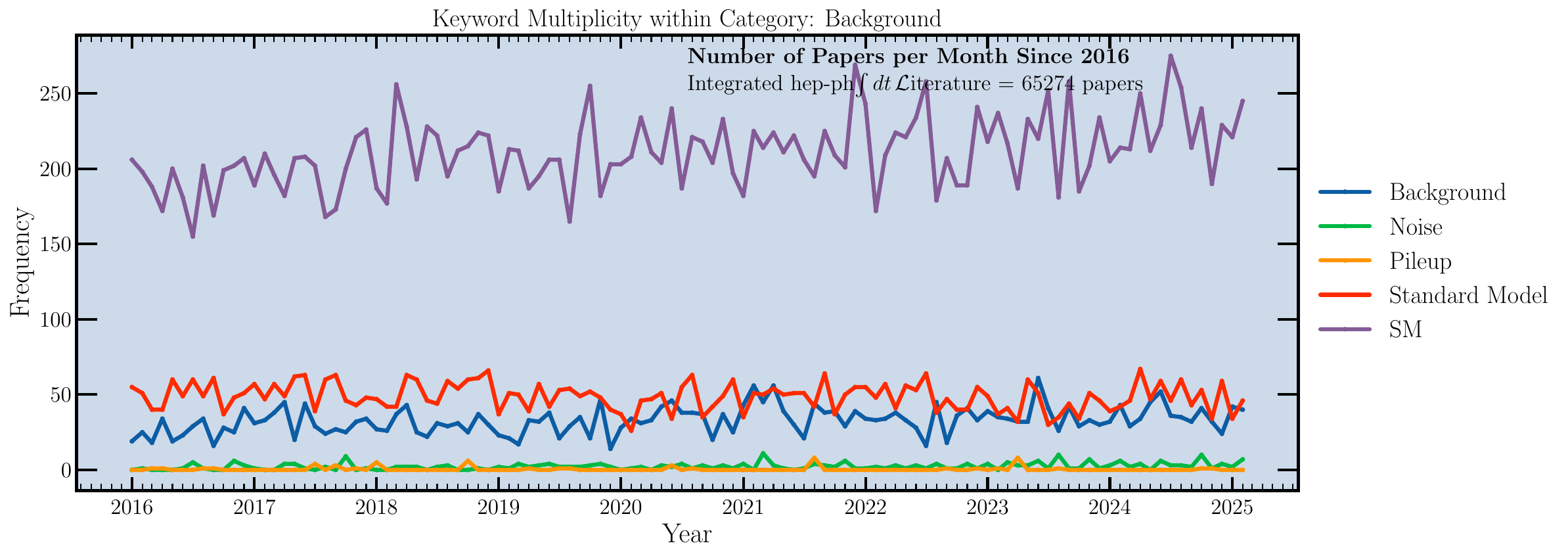}%
        \label{fig:background}%
    }
    \caption{Observed event counts per month for the (a) ``New Physics'' and (b) ``Background'' categories.}
    \label{fig:ph1}
\end{figure}

\begin{figure}[tbp]
    \centering
    \subfloat[the LHC]{%
        \includegraphics[width=0.99\linewidth]{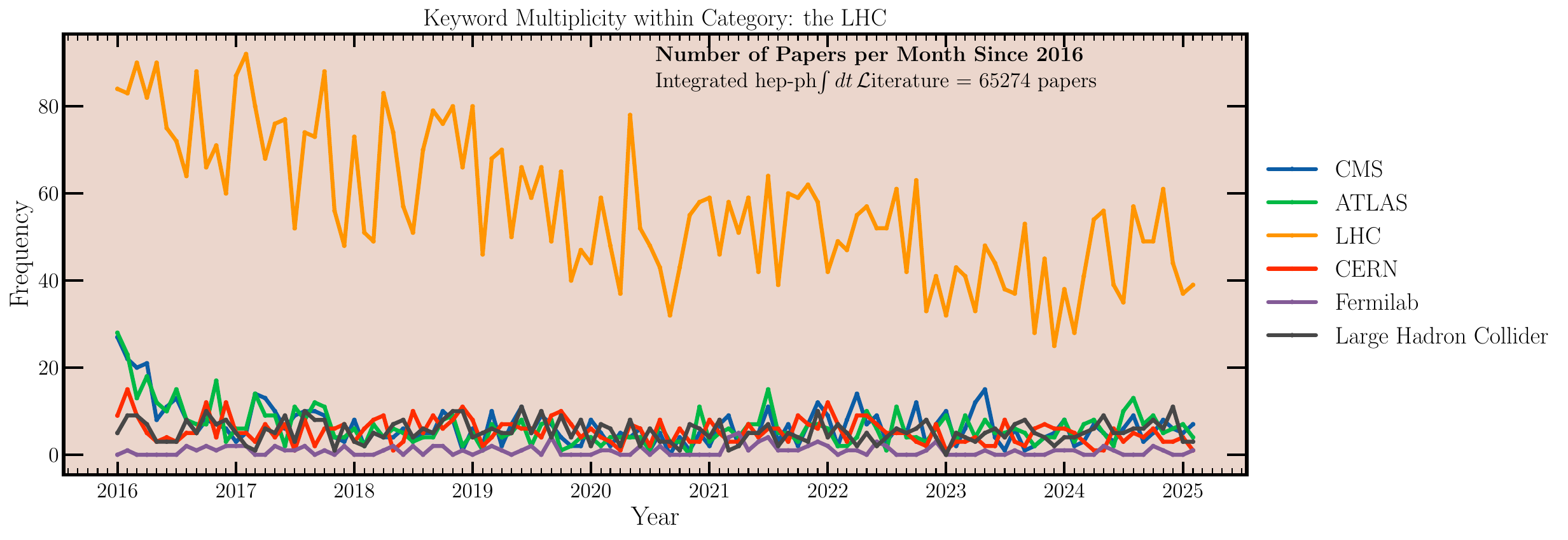}%
        \label{fig:lhc}%
    }\\[0.1cm]
    \subfloat[Future Colliders]{%
        \includegraphics[width=0.99\linewidth]{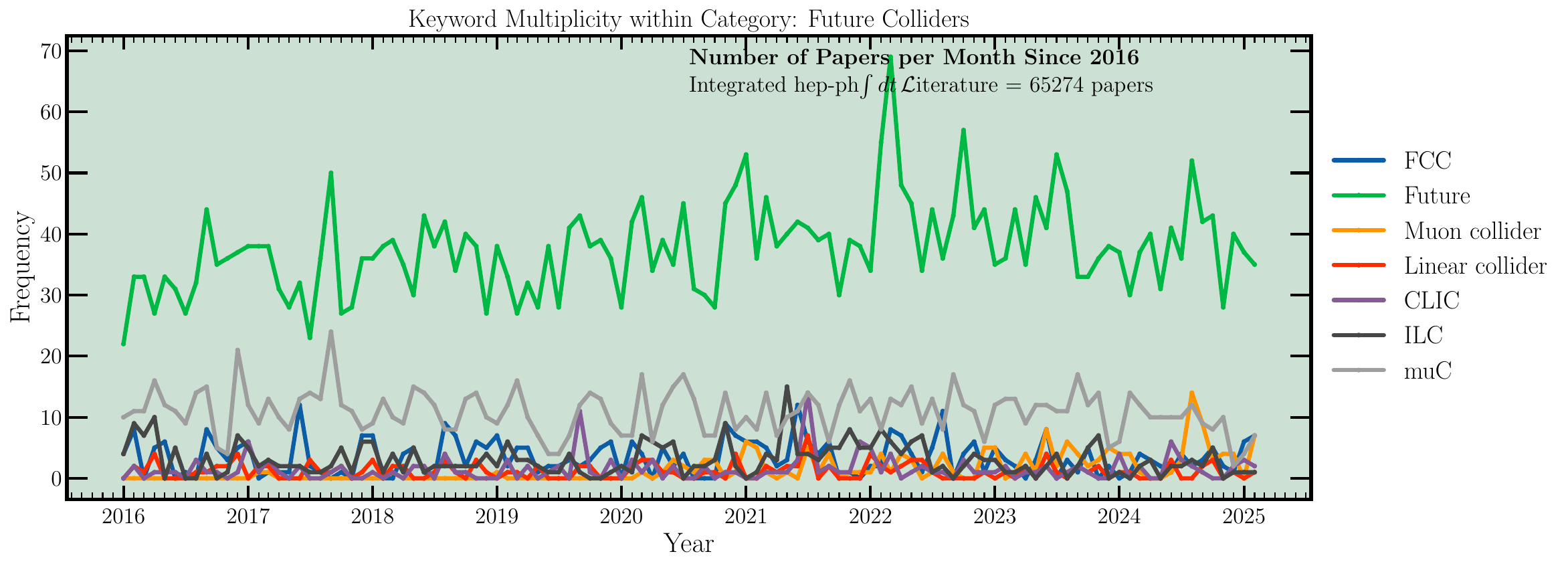}%
        \label{fig:futurecolliders}%
    }
    \caption{Observed event counts per month for the (a) ``the LHC'' and (b) ``Future Collider'' categories.}
    \label{fig:ph2}
\end{figure}

In \Figs{ph1}{ph2}, we break down the distributions of each category per individual keyword. 
This allows us to gain more insight about trends.
We can see in \Fig{newphysics} that of all of the ``New Physics'' category, nearly half of it is ``Dark''.
This is presumably dark matter~\cite{HOLDOM1986196} or dark energy, but it is impossible to know for sure without actually reading the relevant papers, which is completely impossible.
Given that the next-two-most common entries are ``Axions''~\cite{Adams:2022pbo} and ``ALP'', it is likely that we have found our new physics candidate.

We can see in \Fig{lhc} that ``ATLAS'' and ``CMS'' are mentioned in abstracts approximately just as often as eachother.
This is despite the fact that ATLAS has only $\sim$ 3000 authors while CMS has $\sim$5000.
ATLASt, we finally have numerical evidence as to which collaboration is more popular-per-person amongst the \hepph\, community. 
However, both are dwarfed by the total number of ``LHC'' mentions, which has the implication that the \hepph\, community finds the 27 kilometers of tunnels in the LHC far more interesting than either detector.\footnote{It is possible the entire difference is made up by ``LHCb'', as ``LHC`` is a substring of ``LHCb''.}
We can also see in \Fig{futurecolliders} that the ``Muon Collider'' begins to eek out in recent years, nearly tying ``CMS'' and ``ATLAS''.

The most common single keyword, seen in ``\Fig{futurecolliders}'', is ``SM''.
This should be no surprise, as the ``SM'' is typically the most dominant  background.
This also highlights a source of systematic uncertainty in our analysis, induced by IRC-safety: ``sm'' just happens to be a common combination of letters, and IRC-safety removed spaces.
However, as this primarily affects the ``Background'' category, we will leave these in as a conservative estimate of significances.

\begin{figure}
    \centering
    \subfloat[$\hepph$]{%
    \includegraphics[width=1.00\linewidth]{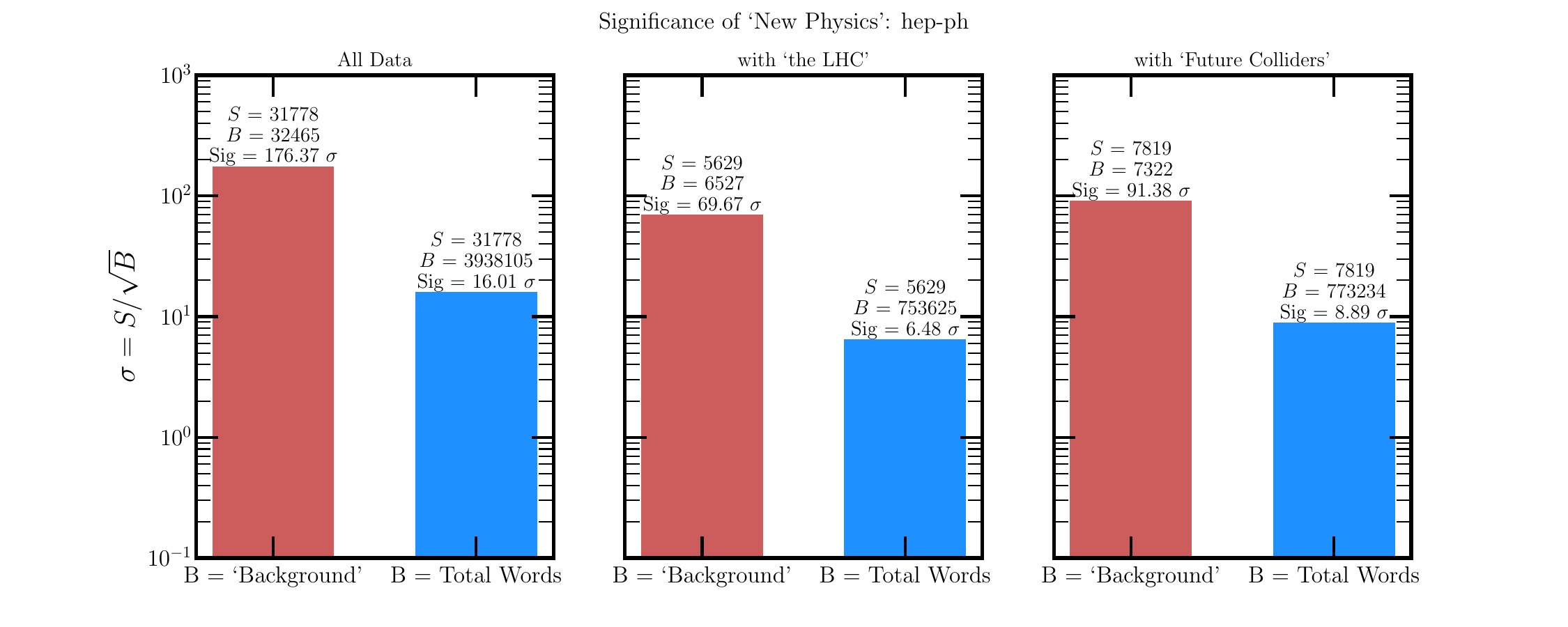}}
    \\
    \subfloat[$\hepex$]{%
    \includegraphics[width=1.00\linewidth]{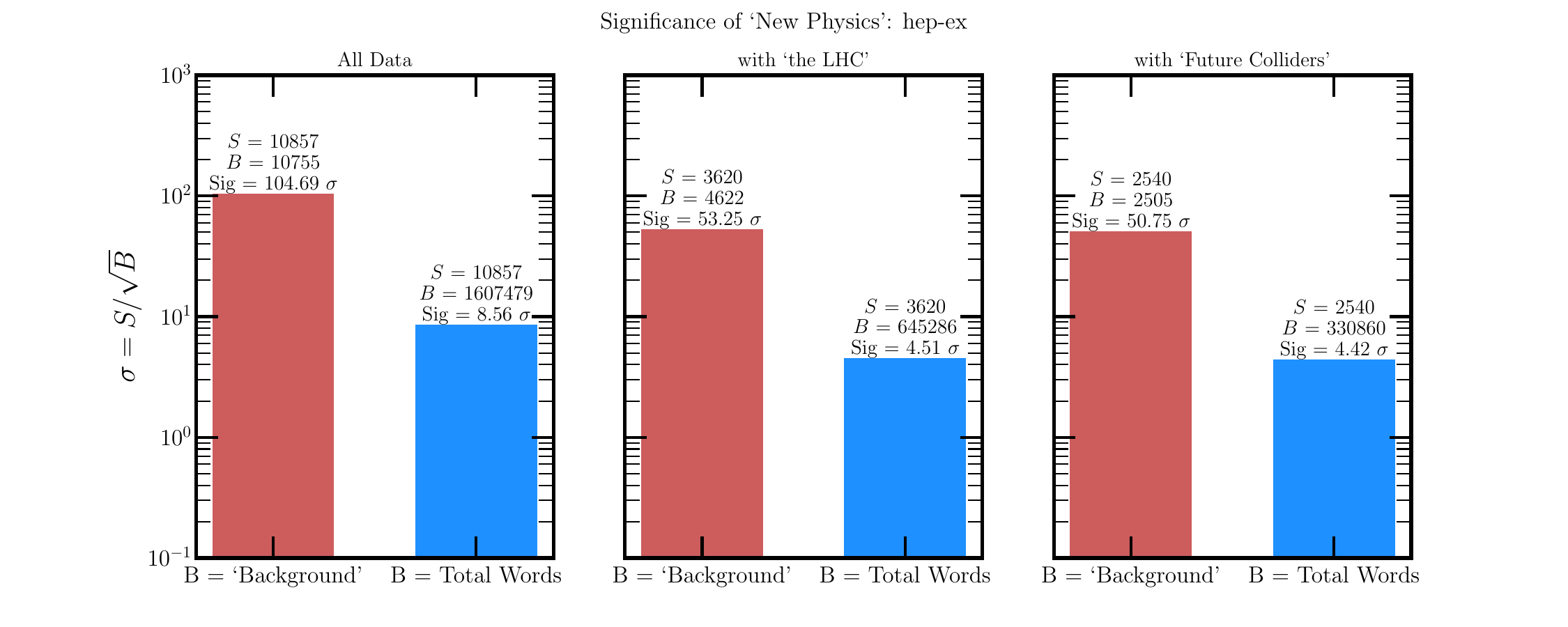}}
    \caption{The estimated significance of searching for ``New Physics'' over the ``Background''s : the naive background consisting of all words (red), and the less-naive background consisting of all ``Background'' keywords (blue). (Left) no condition (Middle) Conditioned on the appearance of an ``LHC'' keyword (Right) Conditioned on the appearance of a ``Future Collider'' keyword. Note the log scale. (a) the main \hepph\, channel (b) the control \hepex\, channel.}
    \label{fig:signifiances}
\end{figure}

Finally, in \Fig{signifiances}, we present the results of our ``New Physics'' search.
In the best case, with no conditioning and with our ``Background'' cut, we achieve a significance of 176$\sigma$.
This is \emph{much} higher than the \hepex\, baseline, which is only 105$\sigma$.
That is, given that ordinary experimental analysis has not claimed to see any new physics and treating 105$\sigma$ as a baseline, the \emph{relative} $p$-value of our discovery is:
\begin{align}
    p &= \frac{p_{175}}{p_{125}} \nonumber \\
    &= \frac{\frac{1}{\sqrt{2\pi}175}\exp(-175^2/2)}{\frac{1}{\sqrt{2\pi}105}\exp(-105^2/2)} \nonumber\\
    &= 0.6\exp(-9800)
\end{align}
It is \emph{incredibly} unlikely, well beyond $5\sigma$, for this fluctuation to occur under the assumption that there is no ``New Physics''.
It is somewhat surprising that the community has missed this.

Second, we can see the impact of searching for ``New Physics'' with different colliders in the middle and right panels of \Fig{signifiances}.
We see that in theory (or rather, in phenomenology), searching with ``Future Colliders'' is significantly better than with the LHC, but in experimental practice, it makes little difference.
However, given that ``New Physics'' only exists in theory anyways and not experimentally, we will adopt the former viewpoint.

As a final interesting note, the sensitivity of searching for ``New Physics'' is \emph{much} higher when it is not with a collider experiment.
This is in line with the earlier observation that the empty tunnels of the LHC seem to attract more interest than CMS or ATLAS.
Perhaps this, alongside the fact that the most common ``New Physics'' keywords were ``Dark'', ``Axion'', and ``ALP'', might be a hint towards what types of experiments the community may want to invent in.

\section{Conclusions}\label{sec:conclusions}

We have presented a search for ``New Physics'' and other physics signals in an Open Dataset, available on \url{arXiv.org}.
We have found a clear-as-day signal, with a $p$-value $< \exp(-9800)$, indicating that there is indeed ``New Physics'' in the \hepph\, channel of \arxiv. 
These ``New Physics'' signals were hiding in plain sight; it required our specialized analysis to go through this old data to search for the needle in the HAYSTAC~\cite{HAYSTAC:2018rwy}.
This new physics is primarily ``Dark'', and is likely ``Axion'' or ``ALP'' specifically.
While a search with a ``Future Collider'' (at least, in theory) yields a higher significance than with ``the LHC'', this signal is unlikely to be uncovered by collider experiments.
We leave the precise determination of this ``New Physics'' to further study.
Unfortunately, this paper has essentially unblinded the dataset: 
Having read this paper, any future papers you now write are tainted by the knowledge of the keywords in \Tab{keyword_categories} and your conscious decision to include or exclude them, rendering further analysis impossible.
We therefore urge that all physicists cease using the phrase ``New Physics'', or any other keywords in \Tab{keyword_categories}, in all future publications.

Our study highlights an essential methodological lesson for the particle physics community: sometimes the most significant discoveries require looking inward at our literature rather than outward to detectors. Indeed, perhaps the real ``New Physics'' was in the papers we wrote along the way.

\section{Code and Data} \label{sec:code_and_data}

The code and data for this analysis, which can be used to reproduce all plots shown here, can be found at \url{https://github.com/rikab/QuoteNewPhysics}.

The original 126 GB dataset has been preprocessed into a list of randomized abstracts. 
For the \hepph\, dataset, there are 250 abstracts per month, and for the \hepex\, dataset, there are 100 per month.
The size of the entire dataset is approximately 50 MB after preprocessing.
All of the analysis code was made in $< 6$ hours. 
I am extremely ashamed to admit I used ChatGPT to generate a lot of it.
This analysis would have been possible without it, but not by April $1^{\rm st}$.

This study refers to well over 25000 papers. I cannot reasonably hope to cite them all --- I even tried and Overleaf crashed.
I would like to therefore apologize to the $\mathcal{O}(25000)$ authors who will not receive a citation bump on inSPIRE due to this, despite their contributions to the dataset. 
The list of all papers used can be found in the files of the above link.

\section*{Acknowledgements}
I would like to thank Jesse Thaler and Cari Cesarotti for their blessings --- the former for not getting mad at me when I \emph{really} should have been working on my thesis instead, and the latter for authoring the \href{http://caricesarotti.com/n_subjesseness.pdf}{legendary seminal work}~\cite{nsubjesseness} that inspired for this paper.
I would also like to extend a special thanks to you, the reader, for enjoying this in the holiday spirit in which it was intended.

\bibliographystyle{JHEP}
\bibliography{refs, jet_refs}

\providecommand{\href}[2]{#2}\begingroup\raggedright\begin{thebibliography}{10}

\bibitem{Peskin:1997ez}
M.~E. Peskin, \emph{{Beyond the standard model}},  in \emph{{The 1996 European School of High-Energy Physics (formerly CERN / JINR School of Physics)}}, pp.~49--142, 5, 1997, \href{https://arxiv.org/abs/hep-ph/9705479}{{\ttfamily hep-ph/9705479}}.

\bibitem{hepmllivingreview}
{HEP ML Community}, ``{A Living Review of Machine Learning for Particle Physics}.''

\bibitem{ATLAS:2020iwa}
{\scshape ATLAS} collaboration, G.~Aad et~al., \emph{{Dijet resonance search with weak supervision using $\sqrt{s}=13$ TeV $pp$ collisions in the ATLAS detector}}, \href{https://doi.org/10.1103/PhysRevLett.125.131801}{\emph{Phys. Rev. Lett.} {\bfseries 125} (2020) 131801}, [\href{https://arxiv.org/abs/2005.02983}{{\ttfamily 2005.02983}}].

\bibitem{ATLAS:2023azi}
{\scshape ATLAS} collaboration, G.~Aad et~al., \emph{{Anomaly detection search for new resonances decaying into a Higgs boson and a generic new particle $X$ in hadronic final states using $\sqrt{s} = 13$ TeV $pp$ collisions with the ATLAS detector}}, \href{https://doi.org/10.1103/PhysRevD.108.052009}{\emph{Phys. Rev. D} {\bfseries 108} (2023) 052009}, [\href{https://arxiv.org/abs/2306.03637}{{\ttfamily 2306.03637}}].

\bibitem{CMS:2024nsz}
{\scshape CMS} collaboration, V.~Chekhovsky et~al., \emph{{Model-agnostic search for dijet resonances with anomalous jet substructure in proton-proton collisions at $\sqrt{s}$ = 13 TeV}},  \href{https://arxiv.org/abs/2412.03747}{{\ttfamily 2412.03747}}.

\bibitem{ATLAS:2023ixc}
{\scshape ATLAS} collaboration, G.~Aad et~al., \emph{{Search for New Phenomena in Two-Body Invariant Mass Distributions Using Unsupervised Machine Learning for Anomaly Detection at s=13\,\,TeV with the ATLAS Detector}}, \href{https://doi.org/10.1103/PhysRevLett.132.081801}{\emph{Phys. Rev. Lett.} {\bfseries 132} (2024) 081801}, [\href{https://arxiv.org/abs/2307.01612}{{\ttfamily 2307.01612}}].

\bibitem{ATLAS:2025obc}
{\scshape ATLAS} collaboration, G.~Aad et~al., \emph{{Weakly supervised anomaly detection for resonant new physics in the dijet final state using proton-proton collisions at $\sqrt{s}=13$ TeV with the ATLAS detector}},  \href{https://arxiv.org/abs/2502.09770}{{\ttfamily 2502.09770}}.

\bibitem{knapp2020adversarially}
O.~Knapp, G.~Dissertori, O.~Cerri, T.~Q. Nguyen, J.-R. Vlimant and M.~Pierini, \emph{{Adversarially Learned Anomaly Detection on CMS Open Data: re-discovering the top quark}},  \href{https://arxiv.org/abs/2005.01598}{{\ttfamily 2005.01598}}.

\bibitem{Gambhir:2025afb}
R.~Gambhir, R.~Mastandrea, B.~Nachman and J.~Thaler, \emph{{Isolating Unisolated Upsilons with Anomaly Detection in CMS Open Data}},  \href{https://arxiv.org/abs/2502.14036}{{\ttfamily 2502.14036}}.

\bibitem{Cesarotti:2023sje}
C.~Cesarotti and R.~Gambhir, \emph{{The new physics case for beam-dump experiments with accelerated muon beams}}, \href{https://doi.org/10.1007/JHEP05(2024)283}{\emph{JHEP} {\bfseries 05} (2024) 283}, [\href{https://arxiv.org/abs/2310.16110}{{\ttfamily 2310.16110}}].

\bibitem{Collins:2019jip}
J.~H. Collins, K.~Howe and B.~Nachman, \emph{{Extending the search for new resonances with machine learning}}, \href{https://doi.org/10.1103/PhysRevD.99.014038}{\emph{Phys. Rev.} {\bfseries D99} (2019) 014038}, [\href{https://arxiv.org/abs/1902.02634}{{\ttfamily 1902.02634}}].

\bibitem{arxiv_scraper}
``arxiv.py python package.''

\bibitem{Smith:2016vcs}
W.~H. Smith, \emph{{Triggering at the LHC}}, \href{https://doi.org/10.1146/annurev-nucl-102115-044713}{\emph{Ann. Rev. Nucl. Part. Sci.} {\bfseries 66} (2016) 123--141}.

\bibitem{Placakyte:2011az}
{\scshape H1and for the ZEUS} collaboration, R.~Placakyte, \emph{{Parton Distribution Functions}},  in \emph{{31st International Symposium on Physics In Collision}}, 11, 2011, \href{https://arxiv.org/abs/1111.5452}{{\ttfamily 1111.5452}}.

\bibitem{Komiske:2020qhg}
P.~T. Komiske, E.~M. Metodiev and J.~Thaler, \emph{{The Hidden Geometry of Particle Collisions}}, \href{https://doi.org/10.1007/JHEP07(2020)006}{\emph{JHEP} {\bfseries 07} (2020) 006}, [\href{https://arxiv.org/abs/2004.04159}{{\ttfamily 2004.04159}}].

\bibitem{Cesarotti:2020hwb}
C.~Cesarotti and J.~Thaler, \emph{{A Robust Measure of Event Isotropy at Colliders}}, \href{https://doi.org/10.1007/JHEP08(2020)084}{\emph{JHEP} {\bfseries 08} (2020) 084}, [\href{https://arxiv.org/abs/2004.06125}{{\ttfamily 2004.06125}}].

\bibitem{Ba:2023hix}
D.~Ba, A.~S. Dogra, R.~Gambhir, A.~Tasissa and J.~Thaler, \emph{{SHAPER: can you hear the shape of a jet?}}, \href{https://doi.org/10.1007/JHEP06(2023)195}{\emph{JHEP} {\bfseries 06} (2023) 195}, [\href{https://arxiv.org/abs/2302.12266}{{\ttfamily 2302.12266}}].

\bibitem{Gambhir:2024ndc}
R.~Gambhir, A.~J. Larkoski and J.~Thaler, \emph{{SPECTER: efficient evaluation of the spectral EMD}}, \href{https://doi.org/10.1007/JHEP12(2024)219}{\emph{JHEP} {\bfseries 12} (2025) 219}, [\href{https://arxiv.org/abs/2410.05379}{{\ttfamily 2410.05379}}].

\bibitem{CMS:2008xjf}
{\scshape CMS} collaboration, S.~Chatrchyan et~al., \emph{{The CMS Experiment at the CERN LHC}}, \href{https://doi.org/10.1088/1748-0221/3/08/S08004}{\emph{JINST} {\bfseries 3} (2008) S08004}.

\bibitem{ATLAS:2008xda}
{\scshape ATLAS} collaboration, G.~Aad et~al., \emph{{The ATLAS Experiment at the CERN Large Hadron Collider}}, \href{https://doi.org/10.1088/1748-0221/3/08/S08003}{\emph{JINST} {\bfseries 3} (2008) S08003}.

\bibitem{Benedikt:2020ejr}
M.~Benedikt, A.~Blondel, P.~Janot, M.~Mangano and F.~Zimmermann, \emph{{Future Circular Colliders succeeding the LHC}}, \href{https://doi.org/10.1038/s41567-020-0856-2}{\emph{Nature Phys.} {\bfseries 16} (2020) 402--407}.

\bibitem{AlAli:2021let}
H.~Al~Ali et~al., \emph{{The muon Smasher\textquoteright{}s guide}}, \href{https://doi.org/10.1088/1361-6633/ac6678}{\emph{Rept. Prog. Phys.} {\bfseries 85} (2022) 084201}, [\href{https://arxiv.org/abs/2103.14043}{{\ttfamily 2103.14043}}].

\bibitem{Brunner:2022usy}
O.~Brunner et~al., \emph{{The CLIC project}},  \href{https://arxiv.org/abs/2203.09186}{{\ttfamily 2203.09186}}.

\bibitem{Battaglia:2007rk}
M.~Battaglia, \emph{{The International Linear Collider}},  in \emph{{Theoretical Advanced Study Institute in Elementary Particle Physics}: {Exploring New Frontiers Using Colliders and Neutrinos}}, pp.~49--99, 5, 2007, \href{https://arxiv.org/abs/0705.3997}{{\ttfamily 0705.3997}}, \href{https://doi.org/10.1142/9789812819260_0002}{DOI}.

\bibitem{HOLDOM1986196}
B.~Holdom, \emph{Two u(1)'s and $\varepsilon$ charge shifts}, \href{https://doi.org/https://doi.org/10.1016/0370-2693(86)91377-8}{\emph{Physics Letters B} {\bfseries 166} (1986) 196--198}.

\bibitem{Adams:2022pbo}
C.~B. Adams et~al., \emph{{Axion Dark Matter}},  in \emph{{Snowmass 2021}}, 3, 2022, \href{https://arxiv.org/abs/2203.14923}{{\ttfamily 2203.14923}}.

\bibitem{HAYSTAC:2018rwy}
{\scshape HAYSTAC} collaboration, L.~Zhong et~al., \emph{{Results from phase 1 of the HAYSTAC microwave cavity axion experiment}}, \href{https://doi.org/10.1103/PhysRevD.97.092001}{\emph{Phys. Rev. D} {\bfseries 97} (2018) 092001}, [\href{https://arxiv.org/abs/1803.03690}{{\ttfamily 1803.03690}}].

\bibitem{nsubjesseness}
``Identifying jet substructure with $n$-subjeseness.''

\end{thebibliography}\endgroup

\end{document}